\documentstyle[multicol,pre,aps,psfig,tighten]{revtex}
\newcommand{\BEQ}{\begin{equation}}     % Gleichungen Anfang ..
\newcommand{\BEA}{\begin{eqnarray}}
\newcommand{\EEQ}{\end{equation}}       % .. und Ende
\newcommand{\EEA}{\end{eqnarray}}
\begin{document}

\input epsf.sty
%\twocolumn[\hsize\textwidth\columnwidth\hsize\csname %
%@twocolumnfalse\endcsname

%\draft

%\widetext

\title{Comment on "Fluctuation-dissipation relations in the
nonequilibrium critical dynamics of Ising models"}

\author{Michel Pleimling}

%\affiliation{
%Institut f\"ur Theoretische Physik I, Universit\"at Erlangen-N\"urnberg,
%D -- 91058 Erlangen, Germany}

\address{
Institut f\"ur Theoretische Physik I, Universit\"at Erlangen-N\"urnberg,
D -- 91058 Erlangen, Germany}
\maketitle

\begin{abstract}
Recently Mayer et al. [Phys.\ Rev.\ E {\bf 68}, 016116 (2003)] proposed a
new way to compute numerically the fluctuation-dissipation ratios in
nonequilibrium critical systems. Using well-known facts of nonequilibrium critical
dynamics I show that the leading contributions of the quantities they consider
are in fact one-time quantities which are independent of the waiting time.
The ratio of these one-time quantities determines the slope of the straight lines
observed in the fluctuation-dissipation plots of Mayer et al.
\end{abstract}
%\maketitle
%\phantom{.}
%]

\begin{multicols}{2}
\narrowtext
In a recent work \cite{May03} Mayer, Berthier, Garrahan, and Sollich
(MBGS in the following) presented a study of ageing phenomena 
taking place in
nonequilibrium Ising models in one and two space dimensions
after a quench from infinite temperature to the critical point located at $T=T_c$.
MBGS present {\it inter alia}
Monte Carlo simulations at the critical point of the two-dimensional
Ising model. These simulations are aimed at computing
fluctuation-dissipation
ratios $X(t,t_w)$ which are defined by
\begin{equation} \label{eq_x}
X(t,t_w)=T_c \, R_{k=0}(t,t_w)/\frac{\partial\, C_{k=0}(t,t_w)}{\partial \, t_w}.
\end{equation}
where $t$ is the time elapsed
since the quench (called observation time) and $t_w < t$ is the waiting time.
$C_{k=0}(t,t_w)$ and $R_{k=0}(t,t_w)$ 
are the long-wave-limits of the
Fourier transforms of the commonly studied spin-spin-correlation function and of
the conjugate response function \cite{God02,Cug02,com1}.
The quantities $\frac{\partial\,C_{k=0}(t,t_w)}{\partial \, t_w}$
and $R_{k=0}(t,t_w)$ have been investigated
field-theoretically by Calabrese and Gambassi in \cite{Cal02}. From general scaling arguments
they are expected to scale in the ageing limit $t_w \gg 1$, $t-t_w \gg 1$ as \cite{Cal02}
\begin{eqnarray}
\frac{\partial\, C_{k=0}(t,t_w)}{\partial \, t_w} & = & A_{\partial \, C}
 \, (t-t_w)^a \, \left( \frac{t}{t_w} \right)^{\theta}
\, F_{\partial \,C}(t_w/t) \label{eq_ck0} \\
R_{k=0}(t,t_w) & = & A_R \, (t-t_w)^a \, \left( \frac{t}{t_w} \right)^\theta
\, F_R(t_w/t) \label{eq_rk0}
\end{eqnarray}
with $a+1 = \frac{2 - \eta}{z}$ and $\theta=\frac{d}{z} - \frac{\lambda_c}{z}-a$.
Here $d$ is the number of space dimensions, $z$ the dynamical exponent,
$\lambda_c$ the autocorrelation exponent, whereas $\eta$ is the usual
equilibrium critical exponent. The functions $F_{\partial \,C}(v)$ and $F_R(v)$ are
universal with $F_{\partial \,C}(0)=F_R(0)=1$.
Eqs.\ (\ref{eq_ck0}) and (\ref{eq_rk0}) can also be written in the
following form:
\begin{eqnarray}
\frac{\partial\, C_{k=0}(t,t_w)}{\partial \, t_w} & = & t_w^{a} \, f_{\partial \,C}(t/t_w) \label{eq_ck02} \\
R_{k=0}(t,t_w) & = & t_w^a \, f_R(t/t_w) \label{eq_rk02}
\end{eqnarray}
where the scaling functions $f_{\partial \,C}(x)$ and $f_R(x)$ vary as
\begin{equation} \label{eq_fcr}
f_{\partial \,C,R}(x) \sim x^{\theta'}
\end{equation}
for $x\gg1$, i.e., $1\ll t_w\ll t$. Here $\theta' = a + \theta$ is the
well-known initial-slip exponent of the magnetization \cite{Jan89} which for the
critical Ising model takes the
value 0.19
%resp.\ 0.108 in two
%resp.\ three space
in two dimensions.
This power-law behaviour (\ref{eq_fcr}) will be of importance in the
following.

In their simulations MBGS do not have direct access to 
$R_{k=0}(t,t_w)$ and $\frac{\partial\, C_{k=0}(t,t_w)}{\partial \, t_w}$. They
instead investigate integrated quantities:
\begin{equation} \label{eq_G}
G(t,t_w) = \int\limits_{t_w}^t \, du \, \frac{\partial\, C_{k=0}(t,u)}{\partial \, u}
= C_{k=0}(t,t) - C_{k=0}(t,t_w)
\end{equation}
and 
\begin{equation} \label{eq_chi}
\chi_m(t,t_w) = T_c \, \int\limits_{t_w}^t \, du \, R_{k=0}(t,u).
\end{equation}
as these quantities are easily obtained in numerical simulations. Plotting $\chi_m(t,t_w)$
against $G(t,t_w)$ they obtain within the accuracy of their
numerical data straight lines with constant slopes. The value of the slope is 
identified by MGBS with the fluctuation-dissipation ratio $X(t,t_w)$, see Eq.\
(\ref{eq_x}), yielding the
claim that $X(t,t_w)$ is
independent of the waiting time $t_w$. Note that this independence on the waiting
time is not supported by the field-theoretical results \cite{Cal02} which yield a
fluctuation-dissipation ratio (\ref{eq_x}) dependent on $t_w$ at two loops. In addition, MBGS
conclude that their ratio $\chi_m(t,t_w)/G(t,t_w)$ gives the limit value $X^\infty =
\lim\limits_{t_w \longrightarrow \infty} \left( 
\lim\limits_{t \longrightarrow \infty} X(t,t_w) \right)$ for all times $t$.

It is the purpose of this Comment to discuss the integrated quantities
involved in the MBGS analysis of the critical two-dimensional Ising model. 
I shall show that the leading contributions to (\ref{eq_G}) and (\ref{eq_chi}) do in fact
not depend on the waiting time  (these one-time quantities will be called
non-ageing in the following). Furthermore, I shall demonstrate that the constant slope
observed by MBGS is given by the ratio of the waiting time independent quantities.
It is therefore not a direct manifestation of ageing, i.e.\ waiting time dependent, behaviour.

The origin of the leading, waiting time independent term is readily understood by
looking at the integrals (\ref{eq_G}) and (\ref{eq_chi}). Inserting the scaling forms
(\ref{eq_ck02}) and (\ref{eq_rk02}) one obtains: $G(t,t_w) \sim t_w^{a+1} \, f_G(t/t_w)$
and $\chi_m(t,t_w) \sim t_w^{a+1} \, f_\chi(t/t_w)$. However in doing so we did not
pay attention to the conditions of validity of (\ref{eq_ck02}) and (\ref{eq_rk02}).
Indeed, close to the upper integration limit these scaling forms cannot be used, since
there the condition $t \gg t_w \gg 1$ is not fulfilled. One might therefore argue that a
time scale $t^*$ exists such that only for $t_w \lesssim  t^*$ the forms 
(\ref{eq_ck02}) and (\ref{eq_rk02}) hold. As shown in the following, the time integrals in
(\ref{eq_G}) and (\ref{eq_chi}) indeed yield a contribution with a scaling behaviour
which differs from that of an ageing quantity.

The correlation $C_{k=0}(t,t_w)$ of the magnetization is given by \cite{note1}
\begin{equation} \label{eq_C}
C_{k=0}(t,t_w)=N \, \left< \left( \frac{1}{N} \sum\limits_{i=1}^N s_i(t) \right) \, 
\left( \frac{1}{N} \sum\limits_{i=1}^N s_i(t_w) \right> \right)
\end{equation}
where $s_i(t)$ is the value of the spin located at the lattice site $i$ at time $t$.
$N$ is the total number of lattice sites, whereas $\left< \cdots \right>$ indicates an
average over the thermal noise. In $G(t,t_w)$, analyzed by MBGS,
this ageing quantity is subtracted from the one-time 
quantity $C_{k=0}(t,t) = N \, \left< \left( \frac{1}{N} \sum\limits_{i=1}^N s_i(t)
\right)^2 \right>$, see (\ref{eq_G}). This latter quantity
is usually denoted by $M^{(2)}(t)$ in the literature and has been extensively studied
in the context of short-time critical dynamics, see, e.g., \cite{Rit96,Oka97,note2}.
Standard scaling arguments \cite{Oka97} show that immediately after the
quench $C_{k=0}(t,t)$ grows as $t^{(d-2\beta/\nu)/z}$
where $d$ is the number of space dimensions and $\beta$ and $\nu$ are the usual equilibrium
critical exponents. For the two-dimensional Ising model we have $(d-2\beta/\nu)/z 
\approx 0.81$. Therefore $G(t,t_w)$ is composed of a non-ageing part (i.e.\ a part
which does not depend on the waiting time) and of an
ageing part where the first one grows much faster in time than the second one.
Indeed, it follows from the dynamical scaling behaviour \cite{Cal02}
that for later times $C_{k=0}(t,t_w) \sim
t^{\theta'}$ with $\theta' = (d-\lambda_c)/z$. Rigorous arguments \cite{Yeun96}
yield the inequality
$\lambda_c \geq d/2$, thus that $C_{k=0}(t,t_w)$ never grows faster than $t^{d/2z}$.
This is illustrated in Fig.\ 1 where 
I plot $G(t,t_w)$ as a function of $t-t_w$ for two of the waiting times ($t_w=46$, $193$) 
considered by MBGS and compare them to $C_{k=0}(t,t)$ (grey line). 
These data have
been obtained in standard simulations of the two-dimensional Glauber-Ising model
with heat-bath dynamics.
%After a few thousand time steps $G(t,t_w)$ 
%grows with the same power-law behaviour as its leading 
%contribution $C_{k=0}(t,t)$. 
The two dot-dashed lines indicate the two different power laws involved:
the leading contribution $\sim t^{0.81}$ and the subleading, ageing contribution
$\sim t^{0.19}$. Clearly $G(t,t_w)$ increases much faster then expected
for an ageing quantity and rapidly displays a behaviour similar to its leading
contribution $C_{k=0}(t,t)$. 
%(the dot-dashed
%line indicates the expected power-law growth $\sim t^{0.81}$). Clearly, the
%contribution of the ageing part $C_{k=0}(t,t_w)$ is 
%of vanishing importance at later times.
As shown in the inset $C_{k=0}(t,t_w)$ itself indeed increases 
as $(t/t_w)^{0.19}$ (dot-dashed lines), in complete agreement with the general scaling
arguments given above. This power-law behaviour
is already encountered for observation times slightly larger than the waiting time.

{
\begin{figure}[htb]
\centerline{\epsfxsize=3.25in\ \epsfbox{
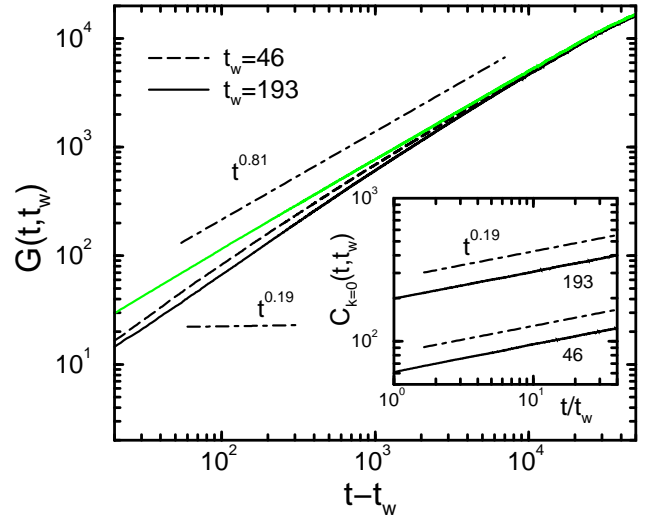}
}
\caption{
$G(t,t_w)$ vs $t-t_w$ for two different waiting times $t_w=46$ and $193$. The grey line
is the leading waiting time independent contribution $C_{k=0}(t,t)$ which grows in time
with a power-law with an exponent $(d-2\beta/\nu)/z \approx 0.81$.
Systems with $300 \times 300$ spins have been simulated, and
all the data shown in the present work have been averaged over 70000 different runs.
The inset shows that the ageing part $C_{k=0}(t,t_w)$ of $G(t,t_w)$ indeed
grows as $t^{0.19}$, as predicted by scaling arguments. 
}
\label{Abb1}
\end{figure}
}

To compute the susceptibility of the magnetization 
a small homogeneous constant field of strength $h$ is switched
on after the waiting time $t_w$ \cite{May03}. The integrated susceptibility is then
given by \cite{note1} 
\begin{equation} \label{eq_Chi}
\chi_m(t,t_w) = \frac{T_c}{N\, h} \langle \sum\limits_{i=1}^N s_i \rangle
= \frac{T_c}{h} m.
\end{equation}
The magnetization $m$ is measured for times $t > t_w$ and depends both 
on $t$ and $t_w$. 
Starting from an uncorrelated initial state, the dynamical correlation length
increases with time, $\xi(t) \sim t^{1/z}$, up to $t_w$. The homogeneous external
field, which is applied for times $t \geq t_w$, 
drives the system away from the critical point towards a new equilibrium
point located at $T=T_c$ and $m=m_{final} > 0$. 
%This is not the case
%when applying a spatially random field in the standard approach. 
%For small fields $m_{final}
%\sim h^{1/\delta}$ where the critical exponent $\delta$ has the value 15 for the
%two-dimensional Ising model. 
This new equilibrium point is reached at finite
times, independently of the waiting time. This is also the case when the
system is already in equilibrium at the critical point and an homogeneous external 
field is then switched on. One therefore expects that the extension of
the correlated regions (and therefore the value of $t_w$)
is only of importance for a short period after the application of the field,
but that at later times the system looses the memory of the value
of $\xi(t_w)$, thus that $\chi_m(t,t_w)$ approaches $\chi_{FC}(t):=\chi_m(t,0)$ for all waiting
times. Here, $\chi_{FC}(t)$ is the field cooling susceptibility.
This is illustrated in Fig.\ 2 where the field strength $h=0.0004$
is the same as that used by MBGS \cite{Ber03}. One indeed observes that the
curves with $t_w \neq 0$ rapidly approach the curve for $t_w=0$. Note that
the plateau reached at longer times is not a finite-size effect, but is due
to the new equilibrium point.
% which for the quantity (\ref{eq_Chi})
%yields the final value $A \, T_c \, h^{1/\delta}/h$ (thin dashed line) with $A = 1.058$
%\cite{Cas00}.
One further remarks from Fig.\ 2 that $\chi_{FC}(t)$ exhibits a power-law behaviour.
The value of the corresponding exponent can be obtained from the standard
dynamical scaling relation for $m$ \cite{Oka97} (with $m(t=0)=0$)
\begin{equation} \label{eq_scal}
m(t,\tau, h) = b^{-\beta/\nu} \, m(b^{-z} \, t, b^{1/\nu} \, \tau, b^{d-\beta/\nu} \, h)
\end{equation}
where $\tau$ is the reduced temperature. In our case $\tau=0$ as the temperature
is fixed at $T_c$ after the quench. Setting $b=t^{1/z}$ one gets
\begin{equation} \label{eq_scal2}
m(t,h)  =  t^{-\beta/\nu z} \, m(1, t^{(d-\beta/\nu)/z} \, h)
\end{equation}
and 
\begin{equation} \label{eq_scal3}
\chi_{FC}(t) = \frac{T_c}{h} m(t,h)  \sim  t^{(d-2\beta/\nu)/ z}
\end{equation}
where the last step $h$ is valid for $t \ll h^{-(d-\beta/\nu)/z}$.
This expected
power-law behaviour is also shown in Fig.\ 2. It is
important to note that $\chi_{FC}(t)$ 
increases with the same exponent
$(d-2\beta/\nu)/ z$ as the leading contribution $C_{k=0}(t,t)$ of the
correlation $G(t,t_w)$, and this already after a few time steps.

{
\begin{figure}[htb]
\centerline{\epsfxsize=3.25in\ \epsfbox{
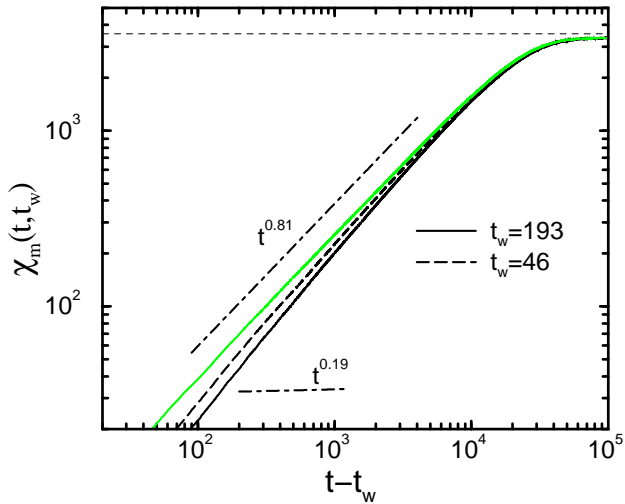}
}
\caption{
The susceptibilty of the magnetization $\chi_m(t,t_w)$ as
function of the time $t-t_w$ elapsed since the switching on of
the homogeneous constant external field. One observes that the
curves for waiting times $t_w \neq 0$ rapidly converge to the curve
obtained for $t_w=0$ where the system is quenched to $T_c$ in presence
of the field. The leading contribution of the susceptibility also
grows with a power-law with an exponent $(d-2\beta/\nu)/z \approx 0.81$.
The thin dashed line indicates the final value of $\chi_m(t,t_w)$
at the new equilibrium point.
}
\label{Abb2}
\end{figure}
}

As the leading non-ageing contributions of both quantities used by MBGS
display the same power-law behaviour, one may wonder whether the straight lines
observed in their fluctuation-dissipation plots, Figs.\ 8 and 10 
in \cite{May03}, are not simply due to this non-ageing parts. 
From the scaling arguments one expects that the ratio $\chi_m(t,t_w)/G(t,t_w)$
is given by $\chi_m(t,t_w)/G(t,t_w) = \chi_{FC}(t)/C_{k=0}(t,t)  
+ O(t^{2\beta/\nu z-\lambda_c/z})$ with $2\beta/\nu z-\lambda_c/z = -0.62$ for the
two-dimensional Ising model.
The ratio $\chi_{FC}(t)/C_{k=0}(t,t)$ is expected to take a constant value $\overline{X}$
already after a few time steps.
This is indeed
the case, as shown in Fig.\ 3. Here I plot
$\chi_{FC}(t)$ as function of $C_{k=0}(t,t)$ and compare the resulting line
with those obtained when plotting $\chi_m(t,t_w)$ as function of
$G(t,t_w)$ for $t_w=46$ and $193$, as done by MBGS. The behaviour of these quantities at the
very first time steps is displayed in the inset. 
It is now obvious
why MBGS obtain straight lines with a constant slope for any value 
of the waiting time $t_w$: the ageing (i.e.\ waiting time dependent)
parts are rapidly suppressed
in this kind of plot
and the slope is then identical to the slope obtained from two 
quantities which do not depend on the waiting time and which furthermore
have the same time dependence. 
%It follows that 
%in the approach of MBGS the fluctuation-dissipation ratio $X_m(t,t_w)$ 
%can not be computed correctly.
%One then also has to conclude that
%their conjecture
%that $X^\infty$ is the same for all observables is not supported
%by any numerical data. 
%
%Finally, let me comment on the value of the slope of the straight line obtained when
%plotting $\chi_m(t,0)$ as a function of $C_{k=0}(t,t)$.
%From the data shown in Fig.\ 3 one obtains that
%the value of the slope is $\approx 0.32$.
%As already noted by MBGS this value is 
%compatible with the limit value of the fluctuation-dissipation ratio
%obtained in the standard approach \cite{God00,Chat03} and in the field-theoretical
%calculations \cite{Cal02}. This raises the interesting question whether
%in nonequilibrium critical systems the limit value $X^\infty =
%\lim\limits_{t_w \longrightarrow \infty} \left(
%\lim\limits_{t \longrightarrow \infty} X(t,t_w) \right)$ of the amplitude ratio
%(\ref{eq_x}) (as studied in \cite{Cal02}) is not simply given by the short time 
%ratio of the amplitudes of
%the one-time quantities $\chi(t,0)$ and $C_{k=0}(t,t)$. Renormalization
%group calculations should be able to clarify this point.

{
\begin{figure}[htb]
\centerline{\epsfxsize=3.25in\ \epsfbox{
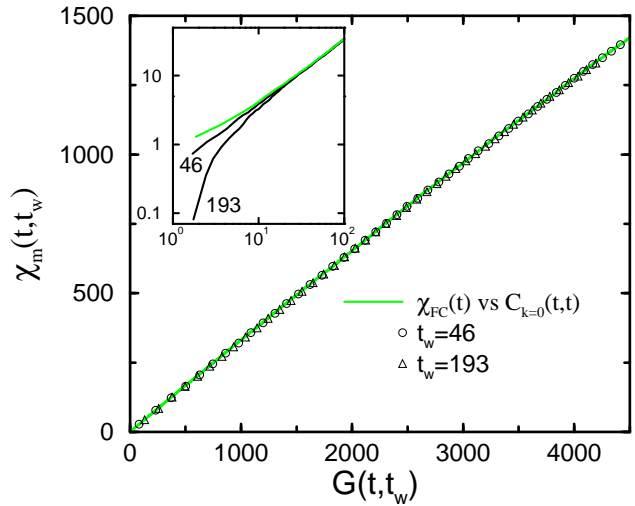}
}
\caption{
Fluctuation-dissipation plot similar to Fig.\ 8 in [1] for waiting times
$t_w=46$ and $193$. Only every 200th data point is shown. The grey line
is the correponding curve obtained from plotting the leading waiting time independent
contribution $\chi_{FC}(t)=\chi_m(t,0)$ of the susceptibility as function of the leading
waiting time independent
contribution $C_{k=0}(t,t)$ of $G(t,t_w)$. Obviously, the slopes of the different curves
are identical after very few time steps 
and are exclusively due to the non-ageing parts of the different quantities, see inset.
}
\label{Abb3}
\end{figure}
}

In conclusion I have shown that the leading terms of the integrated quantities used in \cite{May03}
for the numerical determination of the fluctuation-dissipation ratios
are independent of the waiting time.
I have also shown that 
the leading terms of the correlation (\ref{eq_C}) and
the susceptibility (\ref{eq_Chi}) grow in time with the same power-law.
This explains why MBGS observe in their fluctuation-dissipation plots
straight lines with a constant slope that does neither depend on the waiting time $t_w$ nor on
the observation time $t$.

It is a pleasure to thank A.\ Gambassi for interesting discussions which led to
the present study, and A.\ Gambassi and M.\ Henkel for a critical 
reading of the manuscript. 
This work was supported by the Bayerisch-Franz\"osisches Hochschulzentrum
(BFHZ) and by CINES Montpellier (projet pmn2095). Some simulations have also been done
on the IA32 cluster of the Regionales Rechenzentrum Erlangen (RRZE). I thank G.\ Hager
of the HPC-team of RRZE for technical assistance.

\vspace{-5mm}
%%%%%%%%%%%%%%%%%%%%%%%%%%%%%%%%%%%%%%%%%%%%%%%%%%%%%%%%%%%%%%%%%%%%%%%%%%%%%%%%

%\newpage

\end{multicols}


\begin{references}
\bibitem{com1} MBGS also studied energy related quantities. For simplicity, I will only discuss
the spin-spin correlation function and the conjugate response function. The criticism
formulated in this comment also applies to the energy related quantities.
\bibitem{note1} The quantities here are the same as those
used by MBGS.
\bibitem{note2} I only consider the case where $m(t=0)=0$. In the presence of
an initial magnetization $m(t=0) \neq 0$, the expression for $M^{(2)}(t)$ has
to be slightly modified, see \cite{Rit96}.
\end{references}
\end{document}